\documentclass[%
reprint,
superscriptaddress,
 amsmath,amssymb,
 aps,
]{revtex4-2}

\usepackage{graphicx}
\usepackage{dcolumn}
\usepackage{bm}
\usepackage{amsfonts}
\usepackage{amssymb}
\usepackage{amsmath}
\usepackage{float}
\usepackage{dsfont}
\usepackage[utf8]{inputenc}

\begin{document} 

\title{Microscopic simulations of the coupled dynamics of cavity photons, excitons, and biexcitons}

\author{Hendrik Rose}
\affiliation{Institute for Photonic Quantum Systems (PhoQS), Paderborn University, D-33098 Paderborn, Germany}

\author{Stefan Schumacher}
\affiliation{Institute for Photonic Quantum Systems (PhoQS), Paderborn University, D-33098 Paderborn, Germany}
\affiliation{Department of Physics and Center for Optoelectronics and Photonics Paderborn (CeOPP), Paderborn University, D-33098 Paderborn, Germany}
\affiliation{Wyant College of Optical Sciences, University of Arizona, Tucson, AZ 85721, USA}

\author{Torsten Meier}
\affiliation{Institute for Photonic Quantum Systems (PhoQS), Paderborn University, D-33098 Paderborn, Germany}
\affiliation{Department of Physics and Center for Optoelectronics and Photonics Paderborn (CeOPP), Paderborn University, D-33098 Paderborn, Germany}

\date{\today}

\begin{abstract}
The coherent interaction between quantum light and material excitations in semiconductor nano\-structures is investigated using a fully quantized microscopic approach that incorporates many-body Coulomb correlations. The simulations demonstrate that the quantum dynamics is influenced by biexciton continuum states and is highly sensitive to both the frequency of the cavity mode and the strength of the light-matter coupling.
\end{abstract}

\maketitle

\section{INTRODUCTION}
\label{sec:intro}

Quantum light-matter interactions are important for emerging applications in photonic quantum technologies \cite{Wang2020,Pelucchi2022,Moody2022,Luo2023,Helt:12}. Semiconductor nanostructures and hybrid semiconducting-photonic systems are attractive because their properties can be engineered over a broad range \cite{kira2011semiconductor}. Quantum dots provide simple, discrete energy levels \cite{michler2000quantum,PhysRevB.95.245306,Bracht2021,Jonas2022,Bauch2024,doi:10.1126/sciadv.adw3395}, while larger extended structures such as quantum wells and wires offer stronger but still tunable many-body effects \cite{Weisbuch1992,Deng2010,Rosenberg2018,Luders2023}. The optical properties of these systems have been investigated extensively and very well by microscopic semiclassical approaches, see, e.g.,
Refs.~\cite{Hopfield1958,Sieh1999,10.1088/1464-4266/3/5/201,MEIER2001231,HaugKoch,Kira2006}, which, however, cannot capture quantum features such as fluctuations and entanglement. Fully quantized methods including both Coulomb interactions and quantum-optical effects \cite{PhysRevLett.129.097401,Rose2023,spie2024,528f-7smh} enable more accurate studies of few-photon and nonlinear quantum processes. Microcavity experiments demonstrating strong coupling \cite{PhysRevLett.69.3314,rarity1996microcavities,RevModPhys.71.1591} and clear few-photon polariton behavior \cite{Cuevas2018,acsphotonics.2c01541} highlight the need for models that treat many-body interactions and the light field on the same quantum footing. In this work, we present and analyze numerical simulations obtained from a microscopic and fully quantized model which describes the interactions between microcavity photons, excitons, and biexcitons.

\section{THEORETICAL MODEL}
\label{sec:model}

The system we consider consists of a semiconductor nanostructure embedded in a single-mode optical cavity. Electrons and holes are described by a two-band tight-binding model and the cavity field is treated as a quantized single mode \cite{528f-7smh}. The total Hamiltonian that describes the system's dynamics is given by
\begin{align}
H = H_{S} + H_{LM} + H_{C},
\end{align}
where \(H_{S}\) contains the single-particle electronic and photonic energies, \(H_{LM}\) describes the light-matter coupling that converts photons into electron--hole pairs and vice versa, and \(H_{C}\) accounts for many-body Coulomb interactions among the photoexcited electrons and holes.

The dynamics is formulated in the Heisenberg picture. Expectation values of normal-ordered operators form the basic dynamical variables and their coupled equations of motion follow directly from the Hamiltonian \cite{kira2011semiconductor,Rose2023,528f-7smh}. Here, we assume that the cavity mode is initialized in a two-photon Fock-state and that the electronic system is initially in its ground state. This results in a hierarchy of coupled operator equations which closes naturally as two photons cannot generate more than two electron-hole pairs. Thus no truncation is required and a finite set of equations of motion describes the system dynamics exactly. A coherence-based reduction scheme further simplifies the equations and reduces the numerical effort tremendously as only two-photon coherences, photon-exciton coherences, and biexcitonic coherences are required to capture the full dynamics of the system \cite{528f-7smh}.

The mean photon number is defined as $\langle b^\dagger b \rangle$, where \(b^\dagger\) and \(b\) are photon creation and annihilation operators. Its time evolution follows from the Heisenberg equation,
\begin{align}
\frac{d}{dt} \langle b^\dagger b \rangle
    = \frac{i}{\hbar} \langle [H, b^\dagger b] \rangle ,
\label{mpn}
\end{align}
which couples, as described above, to the dynamics of material, photonic, and mixed material and photonic expectation values. The resulting set of equations of motion can be solved numerically. Further details are provided in Ref.~\citenum{528f-7smh}.

\begin{figure*}
	\centering
		\includegraphics[width=0.91\textwidth]{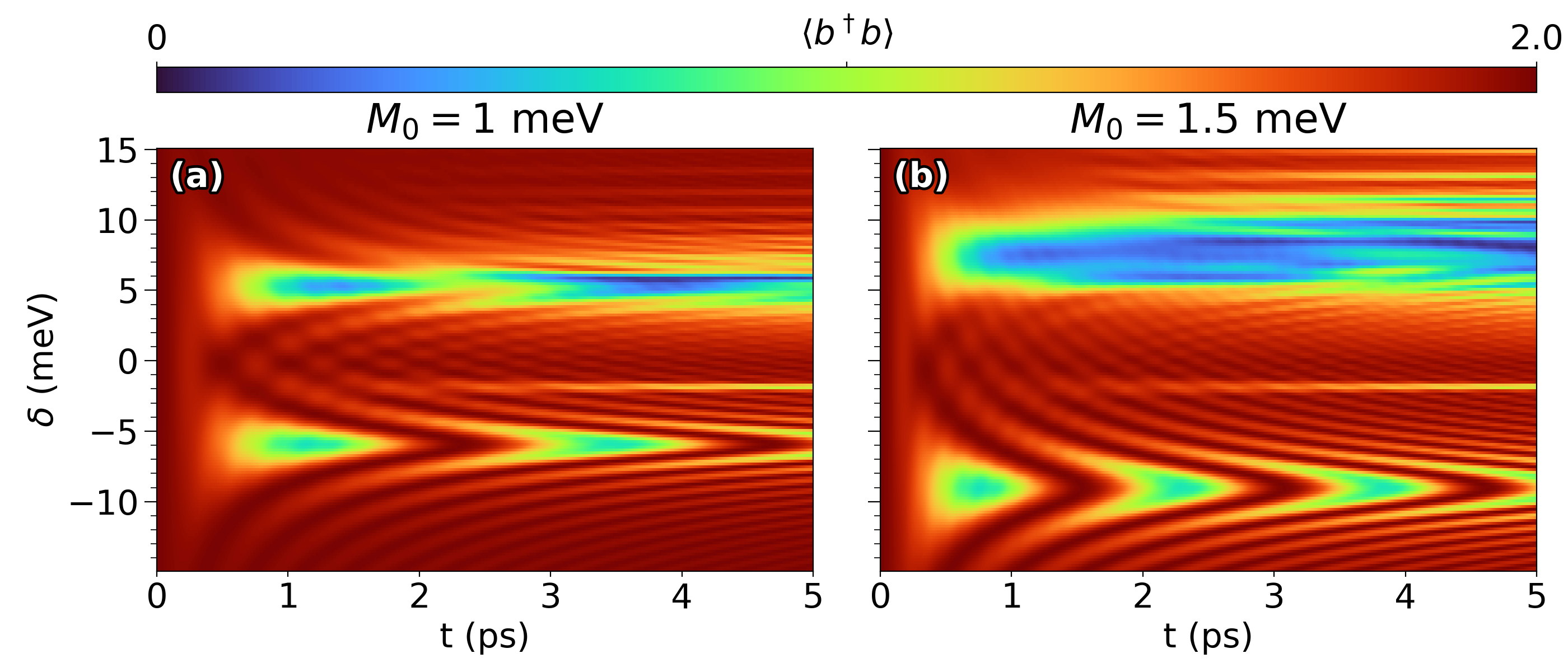}
	\caption{Numerical simulations of Eq.~(\ref{mpn}) using parameters given in the main text.}
	\label{fig1}
\end{figure*}

\section{RESULTS}
\label{sec:sections}

Figure~\ref{fig1} shows the results of simulations of the dynamics of the mean photon number $\langle b^\dagger b\rangle$ for different detunings $\delta$ between the cavity photon energy $E_c$ and the energy of the 1s exciton $E_X$, where $\delta = E_c - E_X$. The parameters used in our simulations are $K = 60$, $U_0 = 20\,\text{meV}$, $J_c = 20\,\text{meV}$, $J_v = 2\,\text{meV}$, $a_0/d = 0.5$, respectively. The light-matter coupling is  $M_0 = 1\,\text{meV}$ in Fig.~\ref{fig1}(a) and $M_0 = 1.5\,\text{meV}$ in Fig.~\ref{fig1}(b); for a detailed description of our model and the parameters we refer to Ref.~\citenum{528f-7smh}.

Our model parameters result in binding energies of $X_b \approx 20.06~\mathrm{meV}$ for the 1s exciton and $XX_b\approx 3.82~\mathrm{meV}$ for the biexciton, respectively. 
Thus an excitation at the band gap would correspond to $\delta \approx 20.06~\mathrm{meV}$
which is above the spectral range around the exciton energy which we consider in Fig.~\ref{fig1}. The weak absorption feature, i.e., transient reduction of the photon number, visible in Figs.~\ref{fig1}(a) and (b) at $\delta \approx -1.9~\mathrm{meV}$, is in good agreement with half the biexciton binding energy, i.e., $- XX_b/2$, and hence provides clear evidence for the presence and the excitation of a bound biexciton state. Most notably, the spectral position is independent on $M_0$. 

Furthermore, we do not observe significant absorption when the cavity mode is in resonance with the 1s exciton, i.e., at $\delta = 0$. Instead Rabi oscillations appear at detunings significantly below the exciton energy. As this shift is not present for reduced models which include only bound states \cite{528f-7smh}, it can be attributed to a strong effective coupling to the continuum of unbound biexciton states. This leads to a significant normal-mode splitting which displaces the resonance. This interpretation is supported by the fact that a larger light-matter coupling $M_0$ results in a larger splitting, cp. Figs.~\ref{fig1}(a) and (b). Despite this shift, the corresponding Rabi oscillation retains excitonic character and its frequency increases with $M_0$. In addition, we identify a spectral region at and above $\delta \approx 5~\mathrm{meV}$ that exhibits a strong reduction of the mean photon number. This feature is consistent with excitation into the continuum of unbound biexciton states. Again, also this behavior cannot be captured by, for example, few models that include only the 1s exciton and the bound biexciton states. Although extended simplified models that incorporate unbound biexcitons phenomenologically may exhibit qualitatively similar resonance shifts, they fail to reproduce the effect quantitatively \cite{528f-7smh}.

\section{CONCLUSIONS}
\label{sec:conc}

We consider a fully quantized and microscopic model to describe the light-matter interaction of a two-photon Fock-state with a semiconductor nanostructure. We present numerical results for the mean photon number, which features prominent shifts of the resonances, Rabi oscillations of mainly excitonic character, and weak absorption of bound biexcitons. These results cannot be captured by simplified models which include only bound exciton and biexciton states. Thus our results highlight that detailed a microscopic analysis is required for an accurate modeling of semiconductor nanostructures interacting with quantum light.
Interesting directions for future work would be to extend the model to multi-mode fields and more general photon statistics, though photon numbers beyond two will require new truncation strategies. 

\acknowledgments   

We are grateful for financial support from the Deutsche Forschungsgemeinschaft (DFG) through the Collaborative Research Center ”Tailored Nonlinear Photonics” TRR 142/3 (project number 231447078, subproject A02).
The authors gratefully acknowledge the computing time made available to them on the high-performance computer Noctua 2 at the NHR Center Paderborn Center for Parallel Computing (PC$^2$). This center is jointly supported by the Federal Ministry of Education and Research and the state governments participating in the NHR (www.nhr-verein.de/unsere-partner). 

\bibliography{main}

\end{document}